\documentstyle[preprint,aps]{revtex}

\begin{document}
\draft
\title {Calculation of Exclusive Cross Sections with the Lorentz Integral
Transform Method}
\author{Alessandro La Piana$^1$ and Winfried Leidemann$^{1,2}$}
\address{
$^{1}$Dipartimento di Fisica, Universit\`a di Trento, I-38050 Povo, Italy;\\
$^{2}$Istituto Nazionale di Fisica Nucleare, Gruppo collegato di Trento.
}

\date{\today}
\maketitle

\begin{abstract}
The longitudinal structure function of the $d(e,e'p)$ exclusive cross
section is calculated with the Lorentz integral transform method. In this 
approach final state interaction is fully taken into account, but without 
using a final state wave function. Cross sections are obtained via the 
inversion of the transform. It is shown that the inversion results are very 
stable. The comparison to a conventional calculation with an explicit $np$ 
final state wave function shows that the obtained results are also very 
precise. Thus the method opens up the possibility to obtain exclusive cross 
sections for reactions with more than two particles, where it is generally 
very difficult to calculate the exact final state wave function.

\end{abstract}

\pacs{ 21.45.+v, 25.30.Fj, 02.30.Qy \\
Keywords: few-body, final state interaction, exclusive reaction, 
Lorentz integral transform}
\section{Introduction}

The study of inclusive and exclusive cross sections in inelastic reactions
is an essential tool in understanding the underlying dynamics of a particle 
system. For systems with more than two constituents a major problem in the 
calculation of such reactions  consists in the exact knowledge of the final 
state wave function in the continuum. Its calculation is far more difficult 
than the corresponding bound state calculation. As a matter of fact, today a 
computation of an intermediate energy continuum wave function is out of reach 
for a system with more than three particles. An exact calculation, however, can
be carried out in an alternative way by using proper integral transforms. They 
allow one to take into account final state interactions (FSI) rigorously 
without using final state wave functions explicitly. In fact in recent years 
the Lorentz integral transform (LIT) method \cite{ELO94} has been successfully
applied to various inclusive breakup cross sections in few-body physics. 
The LIT is expressed in terms of square integrable functions which 
are obtained from inhomogeneous differential equations. The differential
equations can be solved with similar methods as a bound state problem. After 
having calculated the transform one obtains the cross section by inversion
of the transform. A recent overview of the results obtained with the LIT method 
is given in \cite{ELO99}. 

The LIT method has not yet been used for the calculation of exclusive cross 
sections. The only exception is a calculation of the $^4$He spectral function 
\cite{ELO97}, but in this case the 
method proceeds along the same lines as for an inclusive reaction. However, it 
is in principle possible to apply the integral transform method to general 
exclusive reactions as shown by Efros for the Stieltjes transform \cite{Ef85}. 
In the present paper we illustrate the details of the calculation for the 
$d(e',ep)$ reaction with the LIT. It will also allow us to check the
precision of the obtained results by comparing them to those of a 
conventional calculation with an explicit $np$ final state wave function. Such 
a reliability check is necessary, since the use of integral transforms is not 
always unproblematic for the calculation of cross sections. In fact one may 
encounter problems in the inversion of the transform. For example in Ref. 
\cite{ELO92} it was shown for the case of an inclusive reaction that the 
Stieltjes transform is not very appropriated, since it samples contributions 
over a large energy range making the inversion extremely difficult. Later  
the LIT was proposed and it was shown for the test case of the longitudinal
inelastic deuteron form factor in electron scattering that inclusive 
reactions can be safely calculated with this method. Whether the LIT 
is as appropriated also for exclusive reactions cannot be said a priori,
since such a calculation is more complicated. The aim of the present work
is to investigate this question.

The paper is organized as follows. In Sec. \ref{sec:th} we describe how 
cross sections are calculated with the LIT. Details of the calculation 
for the specific reaction under consideration ($d(e,e'p)n$) are given in Sec. 
\ref{sec:d}. The results are illustrated in Sec. \ref{sec:res} and a 
conclusion is drawn in Sec. \ref{sec:con}.

\section{The LIT method}
\label{sec:th}

The starting point of the LIT method \cite{ELO94} is the calculation of an 
integral transform with a Lorentz kernel
\begin{equation}
L(\sigma) = \int d\omega {F(\omega) \over (w-\sigma_R)^2 + \sigma_I^2} \,\,.
\end{equation}
The function $F$ depends on the internal excitation energy $\omega = E_f-E_0$
of a given particle system and contains information about the transition of 
the system from the ground state $|\Psi_0 \rangle$, with energy $E_0$, to the 
final state $|\Psi_f \rangle$, with energy $E_f$, induced by an external probe.
In case of an inclusive reaction $F(\omega)$ denotes the response function
\begin{equation}
F(\omega)=\int d\Psi_f|\langle\Psi_f|\hat O|\Psi_0\rangle|^2 
\delta(E_f-E_0-\omega) 
\,,
\end{equation}
where $\hat O$ is a transition operator which characterizes the specific
process under consideration.

The key point of the LIT method is an evaluation of $L(\sigma)$ without explicit
knowledge of $F(\omega)$. In a second step the function $F$ is obtained 
from the inversion of the transform. The great advantage of the method
lies in the fact that a calculation of the generally very complicated final
state wave function $|\Psi_f\rangle$ can be avoided as will be discussed below. 
On the contrary  a conventional calculation of $F(\omega)$ can only be
carried out with the explicit knowledge of $|\Psi_f\rangle$.

Using completeness one can show that in order to obtain $L(\sigma)$ one 
has to solve the following differential equation
\begin{equation}
(H-E_0 - \sigma^*) |\tilde\Psi_1(\sigma)\rangle = \hat O |\Psi_0\rangle   
\label{DE_incl}
\end{equation}
with
\begin{equation}
\sigma =\sigma_R + i \sigma_I \,\,\,\,\,\,\,\,\,\, \sigma_R,\,\sigma_I > 0 \,,
\end{equation}
where $H=T+V$ is the Hamiltonian of the system under consideration.
Note that the corresponding homogeneous equation has only the trivial solution,
since $H$ has a real eigenvalue spectrum.

The norm of the solution $\tilde\Psi_1(\sigma)$ determines 
the LIT directly:
\begin{equation}
L(\sigma) = \langle \tilde\Psi_1(\sigma)|\tilde\Psi_1(\sigma)\rangle \,.
\end{equation}
Different from a Schr\"odinger equation at positive energies, one has for the
solution of Eq. (\ref{DE_incl})  a 
very simple boundary condition. Due to the localized source at the right hand 
side (rhs) of Eq. (\ref{DE_incl}) $\tilde\Psi_1(\sigma)$ vanishes at large 
distances similar to a bound state wave function. Therefore one can apply the 
same techniques as for the calculation of a bound state wave function. 

The response function $F(\omega)$ serves only for the determination of 
inclusive cross sections. For an exclusive process one needs a more detailed 
information about the transition of the system. In fact one has to be able to 
calculate transition matrix elements of the form 
\begin{equation}
T_{fi}(E_f) = \langle \Psi_f| \hat O| \Psi_0\rangle \,.
\label{Tfi}
\end{equation}
How such a calculation can be carried out with an integral transform was shown 
for the case of the Stieltjes transform \cite{Ef85}.  For the LIT the 
calculation proceeds analogously as outlined in the following (see also 
\cite{La99,Ef99}). 

For simplicity we will consider an exclusive reaction leading to a final state 
with two fragments, but the method can be applied also for channels with more 
than two particles. Besides the correct final state wave
function $|\Psi_f\rangle$ we also introduce the corresponding plane wave
\begin{equation}
\Phi^{\rm PW}(\vec r) = {\cal A} \Psi_1 \Psi_2 {\exp(i\vec k\cdot\vec r)
\over (2\pi)^{3/2}} \,\,,
\end{equation}
where ${\cal A}$ is a proper antisymmetrizer and $\Psi_1$ and $\Psi_2$
are the internal wave functions of the two fragments, while $\vec k$
and $\vec r$ are the usual relative coordinates for momentum and position 
of the two-body system formed by the two fragments. 

In order to calculate the $T$ matrix element one starts from
the Lippmann-Schwinger equation for the final state
\begin{equation}
\langle \Psi_f| = \langle \Phi^{\rm PW}| + \langle \Phi^{\rm PW}| 
\hat V {1 \over E_f + i\epsilon -H} \,\,
\end{equation}
with
\begin{equation}
\hat V = \sum_{i,j} \hat V_{ij}\,,\,\,\,\,\,\, i\in {\cal F}_1 \,,
j \in {\cal F}_2\,,
\end{equation}
where ${\cal F}_1$ and ${\cal F}_2$ contain all particles of the first 
and second fragment, respectively. Inserting the above expression 
in Eq. (\ref{Tfi}) one obtains a sum of two pieces, a Born term
\begin{equation}
T_{fi}^{\rm Born}(E_f) = \langle \Phi^{\rm PW}| \hat O |\Psi_0 \rangle \,,
\end{equation}
and a term depending on FSI
\begin{equation}
T_{fi}^{\rm FSI}(E_f) = 
\langle \Phi^{\rm PW}| \hat V {1 \over E_f + i\epsilon -H}  \hat O | \Psi_0 
\rangle \,.
\end{equation}
The calculation of the Born term is rather simple. 
The main difficulty of the calculation is the determination
of the matrix element depending on FSI.

Using the completeness of the eigenstates $|\Psi(E) \rangle$ of $H$ one can 
rewrite $T_{fi}^{\rm FSI}$ as follows
\begin{eqnarray}
T_{fi}^{\rm FSI}(E_f) &=& 
\int dE \langle \Phi^{\rm PW}| \hat V | \Psi(E) \rangle \langle \Psi(E) |
{1 \over E_f + i\epsilon -H} \hat O | \Psi_0 \rangle \nonumber\\
&=& \int dE F_{fi}(E) {1 \over E_f + i\epsilon -E} \,\,,
\end{eqnarray}
with
\begin{equation}
F_{fi}(E) = \langle \Phi^{\rm PW}| \hat V | \Psi(E) \rangle \langle \Psi(E) |
 \hat O | \Psi_0 \rangle \,.
\end{equation}
One obtains the following formal solution for the FSI term
\begin{equation}
T_{fi}^{\rm FSI}(E_f) = -i \pi F_{fi}(E_f) + {\cal P} \int_{E_0}^{\infty} dE 
{F_{if}(E) \over E_f -E} \,.
\label{restfi}
\end{equation}
For a calculation of $T_{fi}$ one needs to know $F_{fi}$ for any given 
energy. A direct calculation of $F_{fi}$ is of course in general far too
difficult, since one has to determine final state wave functions $\Psi$ for the 
whole eigenvalue spectrum of $H$. On the other hand an indirect calculation
via the LIT is possible. To this end one performs a Lorentz integral
transform of $F_{fi}$, i.e.
\begin{equation}
L(\sigma) = \int_{E_0}^{\infty} dE 
{F_{fi}(E) \over (E- \sigma) (E-\sigma^*)}\,,\,\,\,\,\,\,\,\,\, \sigma_R > E_0
\,.
\end{equation}
Inserting the definition of $F_{fi}$ one finds
\begin{eqnarray}
L(\sigma) &=& \int_{E_0}^{\infty} dE
\langle \Phi^{\rm PW}| \hat V {1 \over H - \sigma}| \Psi(E) \rangle
 \langle \Psi(E)| {1 \over H - \sigma^*} \hat O | \Psi_0 \rangle \nonumber\\
&=&\langle \Phi^{\rm PW}| \hat V {1 \over H - \sigma}
 {1 \over H - \sigma^*} \hat O | \Psi_0 \rangle \nonumber\\
&=& \langle \tilde \Psi_2(\sigma) | \tilde \Psi_1(\sigma) \rangle \,,
\label{scal}
\end{eqnarray}
with
\begin{eqnarray}
|\tilde \Psi_1(\sigma)\rangle &=& {1 \over H - \sigma_R  + i \sigma_I}
\hat O |\Psi_0 \rangle \,,
\label{Psi_incl} \\
|\tilde \Psi_2(\sigma)\rangle  &=& {1 \over H - \sigma_R  + i \sigma_I} 
\hat V |\Phi^{\rm PW} \rangle \,.
\label{Psi_excl}
\end{eqnarray}
It is evident that (\ref{Psi_incl}) leads essentially to the same differential
equation as for the inclusive process: 
\begin{equation}
(H-\sigma_R + i \sigma_I) |\tilde\Psi_1(\sigma)\rangle = 
\hat O |\Psi_0 \rangle \,.
\label{DE_incl2}
\end{equation}
From Eq. (\ref{Psi_excl}) one obtains 
\begin{equation}
(H-\sigma_R + i \sigma_I) |\tilde\Psi_2(\sigma)\rangle = 
\hat V |\Phi^{\rm PW}\rangle \,,
\label{DE_excl}
\end{equation}
which is similar to Eq. (\ref{DE_incl2}),
but for a different source term on the rhs. For a finite range 
potential one has also in this case a vanishing source term for large 
distances. This guarantees also for $\tilde \Psi_2$ an asymptotic boundary 
condition similar to a ground state problem. If $\hat V$ contains  
also the Coulomb potential one cannot proceed exactly in the same way as 
shown here. In this case one has to start from a modified Lippmann-Schwinger
equation, where Coulomb wave function are taken into account \cite{GoW64,Ef99}. 

As shown by Efros \cite{Ef85,Ef99} the integral transform method can be extended
to exclusive processes with more than two fragments. In principle one obtains
equations similar to Eqs. (\ref{scal},\ref{DE_incl2},\ref{DE_excl}). However, 
one cannot guarantee, as in the two fragment case, that the potential $\hat V$ 
vanishes asymptotically, since two of the fragments could
remain close to each other. Therefore it is necessary to choose a different 
solution to the problem. In fact one can rewrite $L(\sigma)$ as follows
\begin{equation}
L(\sigma) = \langle \Phi^{\rm PW}|\hat V|\tilde{\tilde \Psi}_1 \rangle \,,
\label{LIT_double}
\end{equation}
where $|\tilde{\tilde \Psi}_1 \rangle$ is obtained from the solution of
the following differential equation
\begin{equation}
(H-\sigma_R - i \sigma_I) |\tilde{\tilde\Psi}_1(\sigma)\rangle = 
|\tilde\Psi_1(\sigma)\rangle  \,.
\label{DE_double}
\end{equation}
Since $|\tilde\Psi_1 \rangle$ vanishes at large distances, one can again use
bound state methods for the solution of this differential equation. 

There is another possibility to 
calculate the LIT \cite{Ef99}. Starting from the following identity 
\begin{equation}
{1 \over (H-\sigma)(H-\sigma^*)} = {1 \over 2i\sigma_I}
\Bigl( {1 \over H-\sigma} - {1 \over H-\sigma^*} \Bigr) 
\label{DE_double2}
\end{equation}
and defining 
\begin{equation}
|\tilde\Psi_{1}'(\sigma)\rangle  = 
{1 \over H-\sigma_R-i\sigma_I} \hat O|\Psi_0\rangle \,,
\end{equation}
one gets for the transform
\begin{equation}
L(\sigma) = {1 \over 2i\sigma_I} 
\Bigl(\langle \Phi^{\rm PW}|\hat V|\tilde \Psi_{1}'(\sigma)\rangle -
      \langle \Phi^{\rm PW}|\hat V|\tilde \Psi_1(\sigma)\rangle \Bigr)\,.
\label{LIT_double2}
\end{equation} 
It is seen that one has to solve only one type of differential equation. In 
fact for the solution of $|\tilde \Psi'_1(\sigma)\rangle$ it is sufficient to 
solve Eq. (\ref{DE_incl2}) a second time replacing $\sigma^*$ by $\sigma$.

Which of the two approaches for a nonvanishing source term is more appropriate
depends also on the possibility to obtain a precise numerical solution.
Small errors in the solution of Eq. (\ref{DE_incl2}) could lead in both cases
to larger errors for the calculation of the LIT. Therefore we will also
consider in Sec. \ref{sec:res} these additional possibilities to calculate 
the LIT for our realistic test case of the electromagnetic deuteron breakup.

\section{exclusive deuteron breakup}
\label{sec:d}
The exclusive deuteron breakup $d(e,e'p)n$ is governed by
the four structure functions $f_L$, $f_T$, $f_{LT}$, and $f_{TT}$.
In the following we will only consider the longitudinal 
$f_L$. We use the same notation as in Ref. \cite{ArF79}, where
the structure functions are calculated in the final $np$ c.m. system
with $f_L=f_L(E_{np},|\vec q_{c.m.}|,\theta)$. In this system $\vec q_{c.m.}$
denotes the momentum transfer, $E_{np}$ the relative $np$ energy, and
the relative $np$ momentum $\vec k$ has the angle $\theta$ with respect to 
$\hat q_{c.m.}$. The structure functions are expressed in terms of the 
transition matrix $T_{{\rm S}m\mu m_d}$ for the process
$e + d \rightarrow e' + np$. The quantum numbers S and $m$ denote the spin 
and spin projection of the outgoing $np$ pair with respect to $\hat k$, 
$\mu$ characterizes the transition operator $\hat O$, and $m_d$ is the 
projection of the deuteron spin with respect to $\hat q_{c.m.}$.
For the longitudinal structure function, i.e. $\mu=0$, one has  
the following transition operator
\begin{equation}
\hat O = \sum_j G_{E,j}(q_\mu^2) \exp({i\vec q_{c.m.} \cdot \vec r_j}) \,,
\label{op}
\end{equation}
where $G_{E,j}$ denotes the electric form factor of the j-th nucleon
with $q_\mu^2$ being the four-momentum transfer squared. Here we use
the electric dipole form factor for the proton, while the neutron electric
form factor is set to zero.
It is evident that the above operator does not affect the spin, i.e. $S=1$,
and one gets \cite{ArF79}
\begin{equation}
f_L(E_{np}, q_{c.m.}, \theta) = \sum_{mm_d} T_{1m0m_d} T_{1m0m_d}^*   \,,
\end{equation}
where
\begin{equation}
T_{1m0m_d} = C \langle 1m| \hat O| m_d\rangle
\end{equation}
with
\begin{equation}
C = -(2\pi)^{3 \over 2} \sqrt{kM\alpha \over 4\pi} \,, 
\end{equation}
where $M$ denotes the nucleon mass and $\alpha$ is the fine structure constant.
Like $f_L$ the transition matrix depends on $E_{np}$, $q_{c.m.}$, and $\theta$.
In the following we suppress the dependences on $q_{c.m.}$ and $\theta$,
while the dependence on $E_{np}$ is made explicit in most cases.

The aim of the present work is a test of the LIT method for the exclusive 
deuteron breakup. Since there is no need to obtain results with a realistic 
potential we choose the semi-realistic TN potential \cite{ELO99} for this 
test. It is a central potential model, which is different for the various 
spin-isospin (ST) channels, i. e. $V(r)=\sum_{\rm ST} V^{\rm ST}(r)$. Since we 
have $S=1$, only $V^{10}$ and $V^{11}$ have to be considered here. In 
addition $V^{11}(r)=0$ for the TN potential, hence FSI effects appear only in 
the channel ST=10.
 
Due to the absence of the tensor force in the TN potential one obtains a 
reduced complexity of the equations to be solved. Nevertheless, as calculations 
with this potential model show (see e.g. \cite{ELO99}) results are
sufficiently realistic in order to serve as test case.

As discussed in Sec. \ref{sec:th} there are two contributions for any $T$-matrix
element, a Born and an FSI term. Considering an $s$-wave deuteron with radial 
wave function $u(r)$ one obtains for the Born term
\begin{equation}
T_{1m0m_d} = (-)^{m_d} \delta_{mm_d} G_{E,p} \sqrt{3kM} Y_{00}(\hat k_{-})
\int_0^\infty dr r u(r) j_0(k_{-}r) \,,
\label{born}
\end{equation}
where $j_0$ denotes the spherical Bessel function of order 0 and
\begin{equation}
\vec k_- = \vec k - {\vec q_{c.m.} \over 2} \,.
\end{equation}
For the second piece, $T^{\rm FSI}$, it is necessary to perform a multipole
decomposition. It is convenient to introduce the following expansions
with projections $M$ with respect to $\hat q_{c.m.}$
\begin{eqnarray}
\tilde \Psi_{1,M} &=& \sum_{j,L} i^L [Y^{[L]}(\hat r) \times 
\chi^{[1]}(\vec\sigma_1,\vec\sigma_2)]^{[j]}_M 
\sqrt{2L+1} \, C^{L1J}_{0MM} \,r^{-1} \tilde \psi^{(1)}_{Lj}(r) \,, \\
\tilde \Psi_{2,M} &=& \sum_{j,l,m_l} i^l  [Y^{[l]}(\hat r) \times 
\chi^{[1]}(\vec\sigma_1,\vec\sigma_2)]^{[j]}_M 
C^{l1j}_{m_lmM} Y^*_{lm_l}(\hat k) r^{-1} \tilde \psi^{(2)}_{lj}(r) \,,
\end{eqnarray}
where $\chi^{[1]}(\vec\sigma_1,\vec\sigma_2)$ denotes the spin wave function 
for a two-nucleon system with $S=1$.
For the rhs of the differential equations (\ref{DE_incl2})
and (\ref{DE_excl}) we perform similar expansions as for $\tilde \Psi_1$
and $\tilde \Psi_2$ leading to
\begin{eqnarray}
\hat O |m_d\rangle &=& \sum_{j,L} i^L  [Y^{[L]}(\hat r) \times 
\chi^{[1]}(\vec\sigma_1,\vec\sigma_2)]^{[j]}_{m_d} 
\sqrt{2L+1} \, C^{L1j}_{0m_dm_d} \,r^{-1} f_{L}(r) \,,\\
f_{L}(r) &=& j_L({q_{c.m.}r \over 2}) u(r)
\end{eqnarray}
and
\begin{eqnarray}
\hat V |\Phi^{\rm PW}_M  \rangle &=& \sum_{j,l,m_l} i^l  [Y^{[l]}(\hat r) 
\times \chi^{[1]}(\vec\sigma_1,\vec\sigma_2)]^j_M 
C^{l1jM}_{m_lmM} Y^*_{lm_l}(\hat k) r^{-1} g_{lj}(r) \,, \\
g_{lj}(r) &=& \sqrt{2 \over \pi} \, r j_l(kr) V_{jl}(r) \,,
\end{eqnarray}
where $V_{jl}$ is the potential for the NN partial wave $^3l_j$.

For $\tilde \Psi_2$ the above multipole decompositions lead to the 
following coupled differential equations in real and imaginary parts
\begin{eqnarray}
\Bigl\{ -{\hbar^2 \over M} \left( {d^2 \over dr^2} - {l(l+1) \over r^2} \right)
+ V_{lj}(r) - \sigma_R \Bigr\} \Re[{\tilde\psi_{lj}^{(2)}}(\sigma,r)]
- \sigma_I \Im[{\tilde\psi_{lj}^{(2)}}(\sigma,r)] &=&
g_{lj}(r) \\
\Bigl\{ -{\hbar^2 \over M} \left( {d^2 \over dr^2} - {l(l+1) \over r^2} \right)
+ V_{lj}(r) - \sigma_R \Bigr\} \Im[{\tilde\psi_{lj}^{(2)}}(\sigma,r)]
+ \sigma_I \Re[{\tilde\psi_{lj}^{(2)}}(\sigma,r)] &=& 0 \,. 
\end{eqnarray}
As mentioned above, because of our potential model we have to consider only the
channel with ST=10, thus $V_{lj}(r)$ can be replaced by $V^{10}(r)$.  
In addition the Pauli principle has to be fulfilled, i.e. S+T+$l$ has to be 
odd. Therefore the differential equation has only to be solved for $l$ even.
For channels with l odd one had to consider $V^{11}$, but as already mentioned
$V^{11}$ is zero in our potential model.
Note that there is no explicit dependence on $j$ in the coupled differential 
equation, thus one has $\tilde\psi_{lj}^{(2)}=\tilde\psi_{lj'}^{(2)}$.

For $\tilde\Psi_1$ one finds very similar equations with the only difference 
that one has to replace $g_{lj}(r)$ by $f_{L}(r)$ on the rhs.
Also here we have $\tilde\psi_{Lj}^{(1)}=\tilde\psi_{Lj'}^{(1)}$.
 
We solve the differential equation by adding an additional homogeneous
equation determining the source terms on the rhs. In this way 
we obtain a coupled homogeneous differential equation system. The numerical
solution leads to very precise results as shown in Ref. \cite{La99}. 

With the solutions for $\tilde \psi^{(1)}_{Lj}$ and $\tilde \psi^{(2)}_{lj}$
one obtains for the scalar product (\ref{scal})
\begin{equation}
\langle \tilde \Psi_2| \tilde \Psi_1\rangle =
\sum_{mm_dlm_l} \sqrt{2l+1} \, Y_{lm_l}(\hat k) 
\sum_j C^{l1j}_{m_lmm_d} C^{l1j}_{0m_dm_d} \tilde R_{lj}(\sigma)
\end{equation}
with
\begin{equation}
\tilde R_{lj}(\sigma) = \int_0^{\infty} dr (\tilde\psi^{(2)}_{lj}(\sigma,r))^*
\tilde\psi^{(1)}_{Lj}(\sigma,r) \delta_{Ll} \,.
\end{equation}

To calculate the FSI contribution to the $T$-matrix elements
one has to invert the LIT 
\begin{equation}
\tilde R_{lj}(\sigma) = \int_{E_0}^{\infty} dE {R_{lj}(E) \over
(E - \sigma_R)^2 + \sigma_I^2}\\
\label{LITR}
\end{equation}
in order to obtain the function $R_{lj}(E)$.
The transform can be inverted using the following ansatz 
\begin{equation}
R_{lj}(E) = \sum_{n=1}^N c_{n,lj} \chi_{n,lj}(E,\beta) \,,
\label{sumr}
\end{equation}
where $\chi_{n,lj}$ are given functions with nonlinear parameters $\beta$.
Substituting this expansion into the rhs of Eq. (\ref{LITR}) one obtains
\begin{equation}
\tilde R_{lj}(\sigma) =
\sum_{n=1}^N c_{n,lj} \tilde\chi_{n,lj}(\sigma,\beta) \,,
\label{sumphi}
\end{equation}
where the $\tilde\chi_{n,lj}$ are the Lorentz integral transforms of the basis
functions. The parameters $c_{n,lj}$ and $\beta$ are determined by
fitting the calculated transform $\tilde R_{lj}(\sigma)$
to the above expansion at many $\sigma_R$ points for a fixed $\sigma_I$. The 
number of functions
$N$ plays the role of a regularization parameter and is chosen within 
a stability region, i.e. where the obtained results are stable for a certain
range of $N$ (see also \cite{ELO98,ELO99}).
Here we use the following set of basis functions 
\begin{equation}
\chi_{n,lj}(E, \beta) = E^{l+{1 \over 2}} exp(-{\beta E \over n}) \,.
\end{equation}
For the parametrization of the elastic monopole transition for $l=0$
we include an additional function in the set $\chi$:
\begin{equation}
\chi_{0,0j}(E, \beta) = \delta(E-E_0) \,.
\end{equation}
Thus the sum in Eqs. (\ref{sumr}, \ref{sumphi}) starts in this case with $n=0$ 
instead of $n=1$.

Once the inversion is carried out one can make use of Eq. (\ref{restfi}).
For a specific set of $m$ and $m_d$ one finds the following FSI contribution
to the $T$-matrix
\begin{equation}
T^{\rm FSI}_{1m0m_d}(E_{np}) = 
\sum_{lm_l}Y_{lm_l}(\hat k)\sqrt{2l+1 \over 4\pi}
t^{{\rm FSI},l}_{1m0m_d}(E_{np})
\end{equation}
with
\begin{equation}
t^{{\rm FSI},l}_{1m0m_d}(E_{np}) = C \sqrt{4\pi} G_{E,p}(q_\mu^2)  
\sum_j C^{l1j}_{m_lmm_d} C^{l1j}_{0m_dm_d}
\Bigl( -i \pi R_{lj}(E_{np}) + {\cal P} 
\int_{E_0}^{\infty} dE {R_{lj}(E) \over E_{np} - E}\Bigr)\,.
\label{tfsi}
\end{equation}
For our case without tensor force the following simple relations hold
\begin{eqnarray}
t^{{\rm FSI},l}_{1m0m_d} &=& \delta_{m,m_d} 
t^{{\rm FSI},l}_{1m0m_d}
\label{tl1}
\\
t^{{\rm FSI},l}_{1-10-1} &=& t^{{\rm FSI},l}_{1000}
= t^{{\rm FSI},l}_{1101} \,.
\label{tl2}
\end{eqnarray}

It is worth mentioning that as a byproduct of the calculation one obtains 
also the NN phase shifts from our calculation, i.e. without having  solved the 
Schr\"odinger equation for the scattering state. The ratio of imaginary and 
real parts of a given 
transition matrix element is equal to $tan(\delta)$ (see e.g., Ref. 
\cite{ArF79}). For our simple potential model one obtains
\begin{equation}
\delta_l = atan\Bigl({\Im(t^{{\rm FSI},l}) \over
                 \Re(t^{{\rm FSI},l}) +t^{{\rm Born},l}}\Bigr),
\end{equation}
where $t^{{\rm Born},l}$ is analogously defined as $t^{{\rm FSI},l}$,
and easily evaluated from a multipole decomposition of the Born term in
Eq. (\ref{born}). If one is only interested in the phase shifts 
themselves, one can perform a simpler calculation neglecting
the excitation operator $\hat O$ (see Ref. \cite{Ef85}).
  
\section{discussion of results}
\label{sec:res}
We test the LIT method for exclusive reactions choosing for the electromagnetic
deuteron breakup three different kinematics with rather strong FSI effects: 
(i) in the tail region beyond the quasi-elastic
peak at moderate momentum transfer ($E_{np}=120$ MeV, $q_{c.m.}^2=5$ fm$^{-2}$),
(ii) on the photon line ($E_\gamma=70$ MeV), and (iii) close to the deuteron
breakup threshold ($E_{np}=1$ MeV, $q_{c.m.}^2=2$ fm$^{-2}$). Note for 
kinematics (ii) that there is no longitudinal contribution to the exclusive 
$(e,e'p)$ cross section, but that nevertheless the structure function $f_L$ 
does not vanish. In fact in applying Siegert's theorem the longitudinal matrix 
elements are commonly used in photodisintegration and lead to
the dominant contribution for the electric transitions. 

In order to
have a more detailed comparison between the LIT results and the results of
a conventional calculation we do not simply discuss the final result for $f_L$,
but rather study directly the FSI effect on the various multipole transitions.
This allows us to make a much more precise comparison between the two 
calculations. Because of Eqs. (\ref{tl1},\ref{tl2}) it is sufficient to 
consider $t^{{\rm FSI},l}_{1101}$ in the following.

Before turning to the above mentioned three kinematical cases we first
illustrate results for the transform in a more general way choosing a constant 
$q_{c.m.}^2$ of 5 fm$^{-2}$ and various energies $E_{np}$. In Fig. 1 we show 
$\tilde R_{ll}(\sigma_R,\sigma_I=20\,\,{\rm MeV})$ for $l=0$, 2. Its inversion,
$R_{ll}$, gives a contributes to $t^{{\rm FSI},l}$ (see Eq. (\ref{tfsi})).
For the $l=0$ transforms one has an interesting structure at small $\sigma_R$.
It originates from the rather strong monopole 
transition strength close to the deuteron breakup threshold. It is interesting
to see that the peak becomes more and more pronounced for the case that
also $E_{np}$ moves closer to the threshold region. 
In addition there is a second rather sizable contribution in the low $\sigma_R$ 
range. It arises from the elastic monopole contribution. Therefore one has 
to pay attention in the inversion of $\tilde R_{00}$. One has to check whether
$\sigma_I$ is small
enough to resolve with sufficient precision the elastic contribution from
the threshold contribution. For $l=2$ one has a rather different picture. 
One finds a peak in the quasi-elastic region. Note that for the considered 
momentum of $q_{c.m.}^2$ of 5 fm$^{-2}$ the quasi-elastic peak is situated 
at about $E_{np}=50$ MeV. To find such a 
quasi-elastic peak for $\tilde R_{22}$  is a bit surprising,
since one does not expect there strong FSI effects. On the other hand
FSI should be small not on an absolute scale but compared with the 
corresponding Born term. Furthermore, the real part of the FSI contribution is
difficult to estimate from Fig. 1 because a principle value integral has to
be calculated in this case (see Eq. \ref{tfsi}).

In Fig. 2 we show $R_{ll}(E)$ of kinematics (i) for $l=0$, 2, 4 and 
$\sigma_I=5$ and 20 MeV. It is obtained from the inversion of the corresponding 
$\tilde R_{ll}(\sigma_R,\sigma_I)$ (see Eq. (\ref{LITR})). One sees that six 
basis functions are not
sufficient for the inversion, but for a higher $N$
one obtains a very nice stability of the inversion. Comparing the results
with different $\sigma_I$, one finds a small difference for $l=0$ in the 
threshold region. The differences arise because the monopole contribution has 
a peak at the very threshold which has to be separated from the elastic 
contribution at $E_0=-2.225$ MeV. From the inversion we obtain an elastic 
contribution of about 1.1  fm$^{3 \over 2}$ which is rather sizable compared to 
the inelastic part with a peak height of 0.045  fm$^{3 \over 2}$. 
Therefore is not surprising that the higher 
resolution with $\sigma_I=5$ MeV leads to a somewhat different result. However,
because of the rather high $E_{np}$ of 120 MeV, the difference at
the threshold is rather unimportant. This is confirmed by the results for 
$t^{\rm FSI}$, which are shown in Fig. 3 as function of the number of basis
functions used for the inversion. In fact the agreement among the results with 
$\sigma_I=5$ and 20 MeV is very good. It is seen that one obtains for 
$N \ge 10$ for all considered multipolarities $l$ and for both $\sigma_I$ 
values very similar and stable results.

Also shown in Fig. 3 is the $t^{\rm FSI}$ of a conventional calculation. 
These results are very similar to the LIT results with relative differences
of less than 1\%. Only for the real part of the $l=4$ transition 
the difference is a little bit larger. On the other hand one has also to 
consider that this matrix element is very small. In fact its size is only 
$-2.5$\% of the corresponding $t^{\rm Born}$ matrix element. Thus the 
relative difference for the total matrix element is of the order of $10^{-4}.$
For such a small FSI effect a part of the differences could also be due to a 
not completely exact result of the conventional calculation. Different from 
the LIT method $t^{\rm FSI}$ is not calculated directly, but taken indirectly
from the difference of $t^{\rm total}$ and  $t^{\rm Born}$; here $t^{\rm total}$
corresponds to the transition with the correct $np$ final state wave function
in presence of the potential. The FSI effect  is much more sizable for the two 
other transitions. Taking also here the ratio of $t^{\rm FSI}/t^{Born}$ for the 
real parts, one finds for $l=0$ about $-40$\% and for $l=2$ about $-30$\%.

Results for kinematics (ii) are shown in Figs. 4 and 5. For the $R_{ll}(E)$
of Fig. 4 one finds again nice stabilities of the inversion for a larger 
number of basis functions. Comparing the  $R_{ll}$ with the two different
$\sigma_I$ one has also here differences for the monopole transition and
in addition for $l=4$. The monopole is of course not relevant for this
kinematics on the photon line, since there is no corresponding electric
monopole. On the other hand it is interesting to see whether one is able
to separate the strongly peaked threshold strength from the dominant elastic 
contribution. Due to the lower momentum transfer one obtains an even larger
elastic $R_{ll}$ than for kinematics (i), namely a value of about 14 
fm$^{3 \over 2}$ . The $t^{\rm FSI}$ 
results are shown in Fig. 5. The real parts turn out to be very stable as 
function of number of inversion basis functions. They are also very similar for
both $\sigma_I$. Here we have the following relative FSI effects comparing
with the Born term: $-100$\% ($l=0)$, $-25$\% ($l=2)$, $-1.5$\% ($l=4)$.

In comparison to the conventional calculation one finds in Fig. 5 for all 
the real parts  
only very small differences of less than 1\%.
For the imaginary part of $t^{\rm FSI}$ the picture is a bit different.
The $l=2$ results are very stable and agree with extremely high precision
to the results of the conventional calculation. Also the $l=4$ results are
stable, but they are a few percent larger than found in the conventional 
calculation. However, one should note that the matrix element is very
small and hence the difference of a few percent is not relevant. In fact 
comparing with the above mentioned size of the real part of the total matrix 
element the difference between both calculations is of the order of $10^{-4}$.
The imaginary part of the $l=0$ transition shows a bit less stability with
the number of inversion basis function reflecting also the above mentioned 
problems for the separation of the elastic contribution. On the other hand 
one obtains reliable results for the highest $N$'s. 

For the third kinematics we illustrate the results in Figs. 6 and 7. Here we 
consider only $l=0$ and $l=2$ transitions, since FSI effects do not play any 
role for higher transitions at threshold. In fact the FSI contribution 
is already very small for $l=2$. The inversion results in Fig. 6
are again very stable, except for $l=0$ with $\sigma_I=20$ MeV. Of course,
again it is the problem associated with the elastic contribution
($R_{00}(E_0)$ is about 15 fm$^{3 \over 2}$). Figure 7
shows that one obtains very good results for the real part of $t^{\rm FSI}$
with $\sigma_I=5$ MeV, while there is somewhat less stability for the inversion
results with $\sigma_I=20$ 
MeV. The comparison with the conventional calculation is also here 
satisfactory. There are only
deviations of about 1\%. Again we list the relative FSI effect 
comparing with the Born term: -230\% ($l=0)$, +2.5\% ($l=2)$.
The imaginary parts are somewhat more problematic. For $l=0$ one has 
the already mentioned problem with the elastic contribution combined with
the fact that one needs $R_{00}(E)$ close to threshold ($E_{np}=1$ MeV), but 
with $\sigma_I=5$ MeV one obtains a sufficiently good result as seen from
the comparison to the result of the conventional calculation. Though the 
relative differences to the conventional calculation are rather large for the 
imaginary part of the $l=2$ transition, its value is in principle correct, 
since it is more or less identical to 0. Note that it is about 200 times 
smaller than the already very small real FSI part of the $l=2$ transition. 

We do not show results for the angular distribution of $f_L$. However, from the 
discussion above it should be clear that the two different calculations
lead for $f_L(\theta)$ to relative differences of considerably less than 1\% 
for kinematics (i) and (ii) and of about 1\% for kinematics (iii).

As mentioned in Sec. \ref{sec:th} one has to use a somewhat different method 
for the calculation of the LIT for an exclusive reaction with more than two 
fragments in the final state. Two other possibilities are discussed at the
end of Sec. \ref{sec:th}. In both cases one has to solve different 
differential equations, e.g., (\ref{DE_double}) instead of Eq. (\ref{DE_excl}). 
However, these new methods appear to be numerically more 
problematic. Small errors in $\tilde\Psi_1$, the solution of the differential 
equation (\ref{DE_incl2}), might lead to a much larger error for the 
solution of Eq. (\ref{DE_double}), where $\tilde\Psi_1$ serves as source term 
on the rhs. Also for the determination of the LIT via
Eq. (\ref{LIT_double2}) it is important how precise $\tilde\Psi_1$ and
$\tilde\Psi_1'$ are calculated, since one has to determine the difference
$\langle \Phi^{\rm PW}|\hat V|\tilde \Psi_{1}'(\sigma)\rangle - \langle 
\Phi^{\rm PW}|\hat V|\tilde \Psi_1(\sigma)\rangle$. We are able to study this 
question for the $d(e,e'p)n$ reaction, since we can also use these alternative 
ways of evaluating the LIT. As a matter of fact both alternative methods
lead in our case essentially to the same results with relative differences
smaller than 0.01\%. In Fig. 8 we show for a
few selected cases these new LIT results compared to those obtained with
Eq. (\ref{scal}). On finds relative deviations of the order of 1\%. 
There are larger differences for the kinematics with $E_{np}=1$ 
and 120 MeV beyond a $\sigma_R$ of 150 MeV, but they are rather unimportant, 
since both $\tilde R_{00}$ are very 
small there. In fact $\tilde R_{00}$ crosses zero at about 205 and 190 MeV for 
$E_{np}=1$ and 120 MeV, respectively. Altogether one can say that one does not 
encounter greater numerical problems in evaluating the LIT with these 
alternative ways. Therefore also a calculation of an exclusive reaction to a 
three-body channel should lead to rather reliable results with the LIT method. 
  
\section{conclusion}
\label{sec:con}
We have calculated the longitudinal response of the exclusive $d(e,e'p)$
reaction with the method of the Lorentz integral transform. This method allows
one to include the complete FSI, however, without explicit use of final state 
wave functions. It is the first time that the LIT method is applied to an 
exclusive reaction. In the past only inclusive processes have been studied with 
the LIT. The great success of the method raised the question whether it can 
also be successfully used in exclusive reactions. The results in this work show 
that one obtains a very precise determination of the various transition 
matrix elements. Differences to the conventional calculation are generally 
below 1 \%. Only in the case of a very small FSI effect on the transition 
strength, i.e. a 10$^{-4}$ effect compared to the corresponding Born term, one 
can also obtain somewhat higher differences of a few percent. However, in this 
case differences could, as discussed in Sec. \ref{sec:res}, at least partly be 
due to a small inexactness in the conventional calculation. There is only one 
exception, where one can expect a somewhat larger size of the error of the LIT 
result. This is the case for a transition matrix element in a region with
transition strength from two (or more) rather narrow lying peaks. We had
chosen such a situation with our kinematics (iii), where we have a strong
elastic contribution at about $E=-2.2$ MeV and another strong peak right
above breakup threshold. In such a situation one should try to improve
the resolution of the transform $L(\sigma_R,\sigma_I)$ by choosing a smaller
$\sigma_I$. In fact our results improve significantly from $\sigma_I=20$ MeV
to $\sigma_I=5$ MeV. 

In case of an exclusive reaction with more than two fragments in the final 
state one cannot proceed exactly in the same way as for the breakup
in two fragments. In this case one has to use other ways for the
determination of the LIT. We could show that also these alternative methods
lead to rather precise results. Therefore, in general, we may conclude that 
the LIT method leads to reliable results not only for inclusive, but also for 
 exclusive reactions.  

\section*{Acknowledgment}
We thank V. D. Efros and G. Orlandini for helpful discussions.


\begin{figure}[htb]
\caption
{Lorentz integral transforms $\tilde R_{ll}(\sigma_R,\sigma_I=20$ MeV) for 
$l=0$ (top) and $l=2$ (bottom) at $q_{c.m.}^2=5$ fm$^{-2}$ for various 
$E_{np}$ as shown in the figure.}
\end{figure}

\begin{figure}[htb]
\caption
{Inversion result $R_{ll}(E)$ with $\sigma_I=5$ MeV (left) and $\sigma_I=20$
MeV (right) for $l=0$ (top), $l=2$ (middle), $l=4$ (bottom); as indicated
in the figure curves are shown for a number of inversion basis 
functions $N=6,16,18,20$.} 
\end{figure}

\begin{figure}[htb]
\caption
{Results for real (left) and imaginary (right) parts of 
$t^{{\rm FSI},l}_{1101}$ for kinematics (i)
with $l=0$ (top), $l=2$ (middle), $l=4$ (bottom) as function of the number
of inversion basis functions: diamonds ($\sigma_I=5$ MeV), squares 
($\sigma_I=$20 MeV); 
also shown are the results of a conventional calculation with explicit
final state wave function (full curves) and deviations of $\pm 1$ \% from 
these results (dashed curves).}
\end{figure}

\begin{figure}[htb]
\caption
{As Fig. 2 but for kinematics (ii).}
\end{figure}

\begin{figure}[htb]
\caption
{As Fig. 3 but for kinematics (ii).}
\end{figure}

\begin{figure}[htb]
\caption
{As Fig. 2 but for kinematics (iii) and $l=0$, 2.}
\end{figure}

\begin{figure}[htb]
\caption
{As Fig. 3 but for kinematics (iii) and $l=0$, 2.}
\end{figure}

\begin{figure}[htb]
\caption
{Ratio of Lorentz integral transforms $\tilde R_{00}(\sigma_R,\sigma_I=5$ MeV)
calculated with Eq. (\ref{LIT_double}) (dots) relative to the results of Eq. 
(\ref{scal}) at $q_{c.m.}^2=5$ fm$^{-2}$ for $E_{np}=1$ MeV (top), 120 MeV 
(middle), 200 MeV (bottom).}
\end{figure}

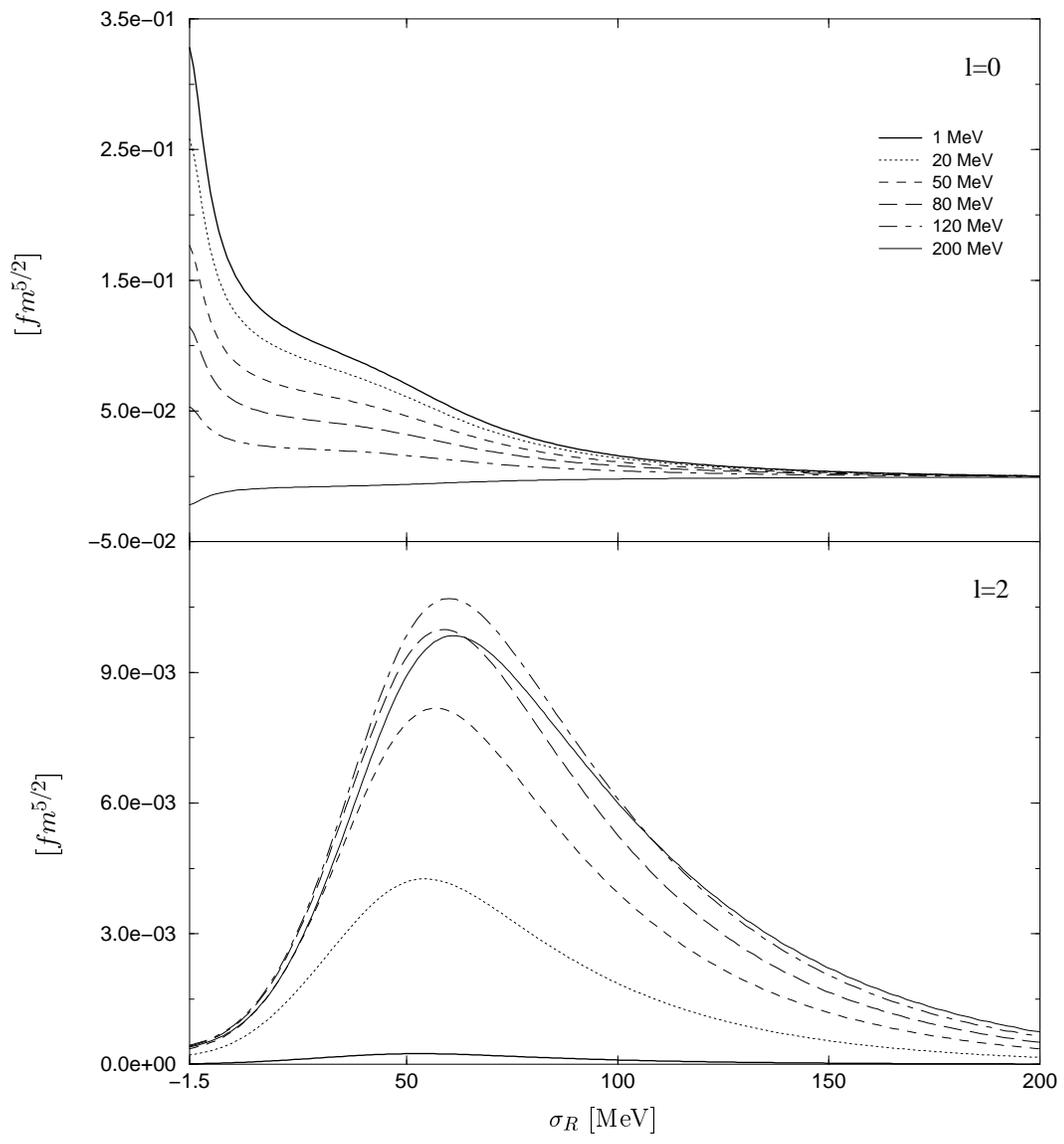
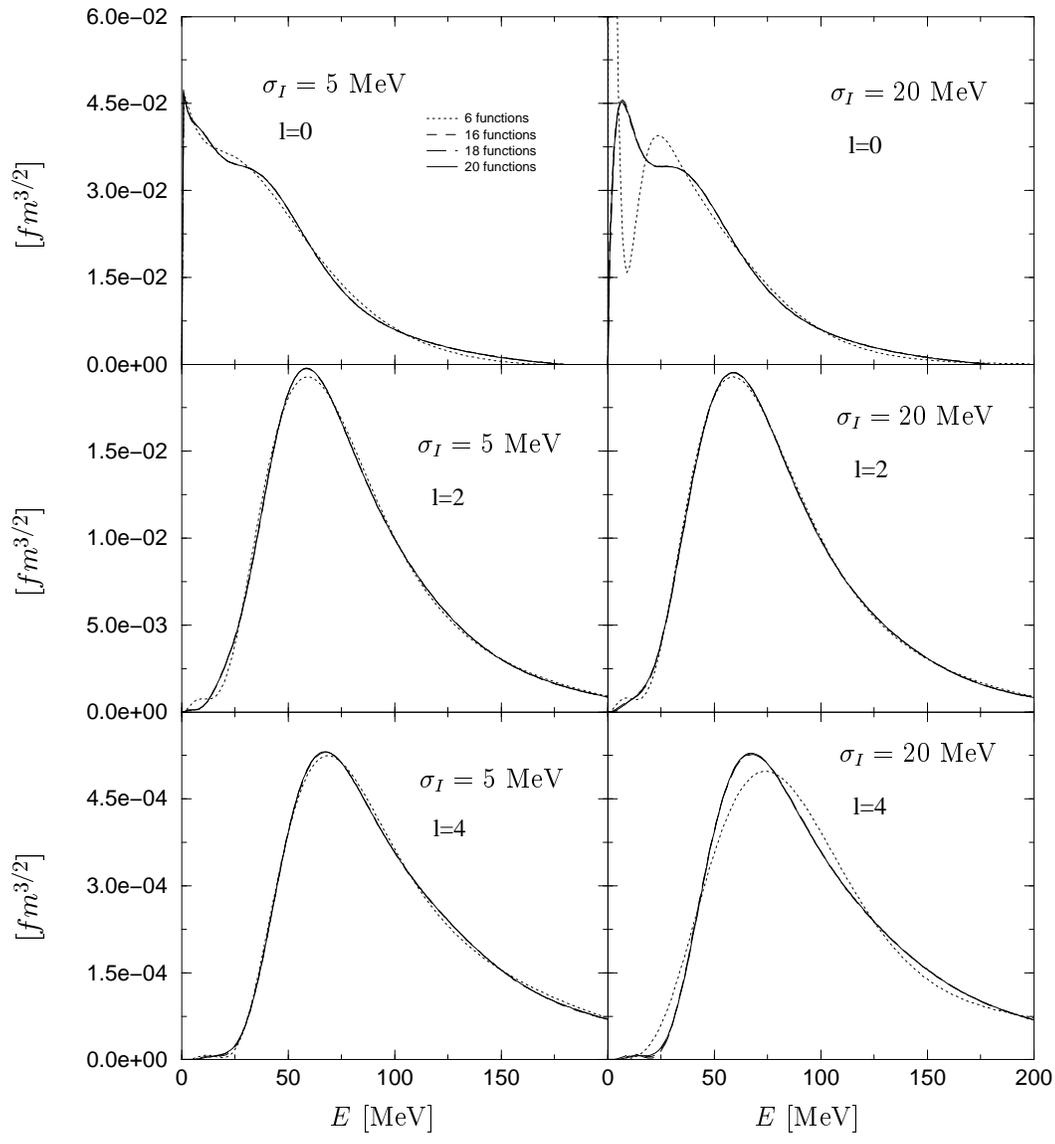
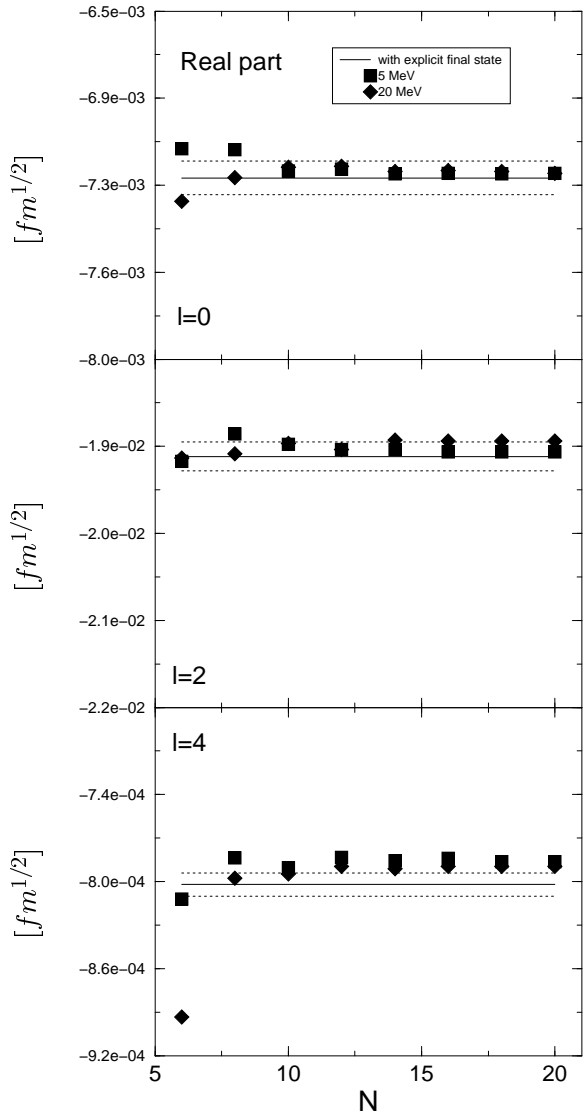
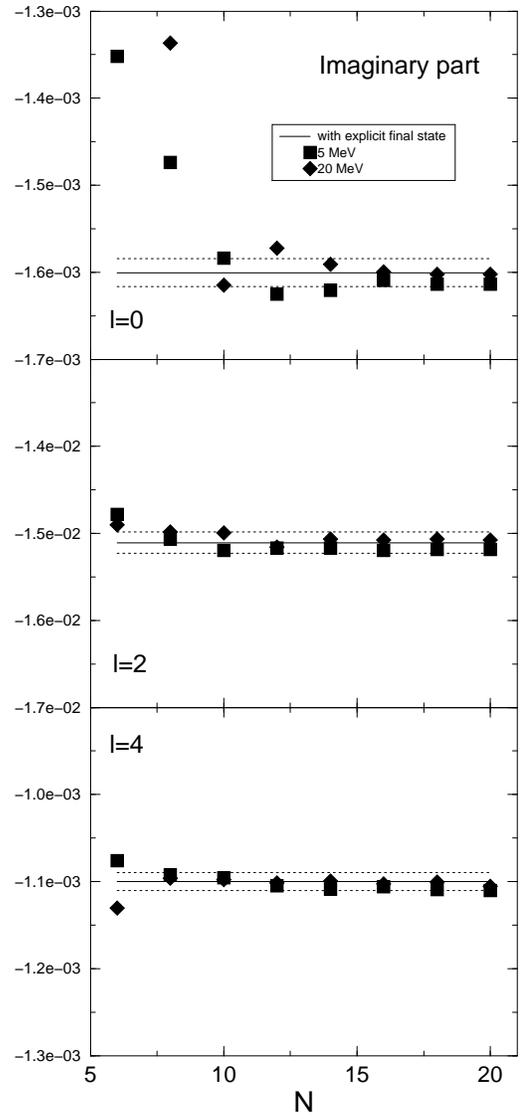
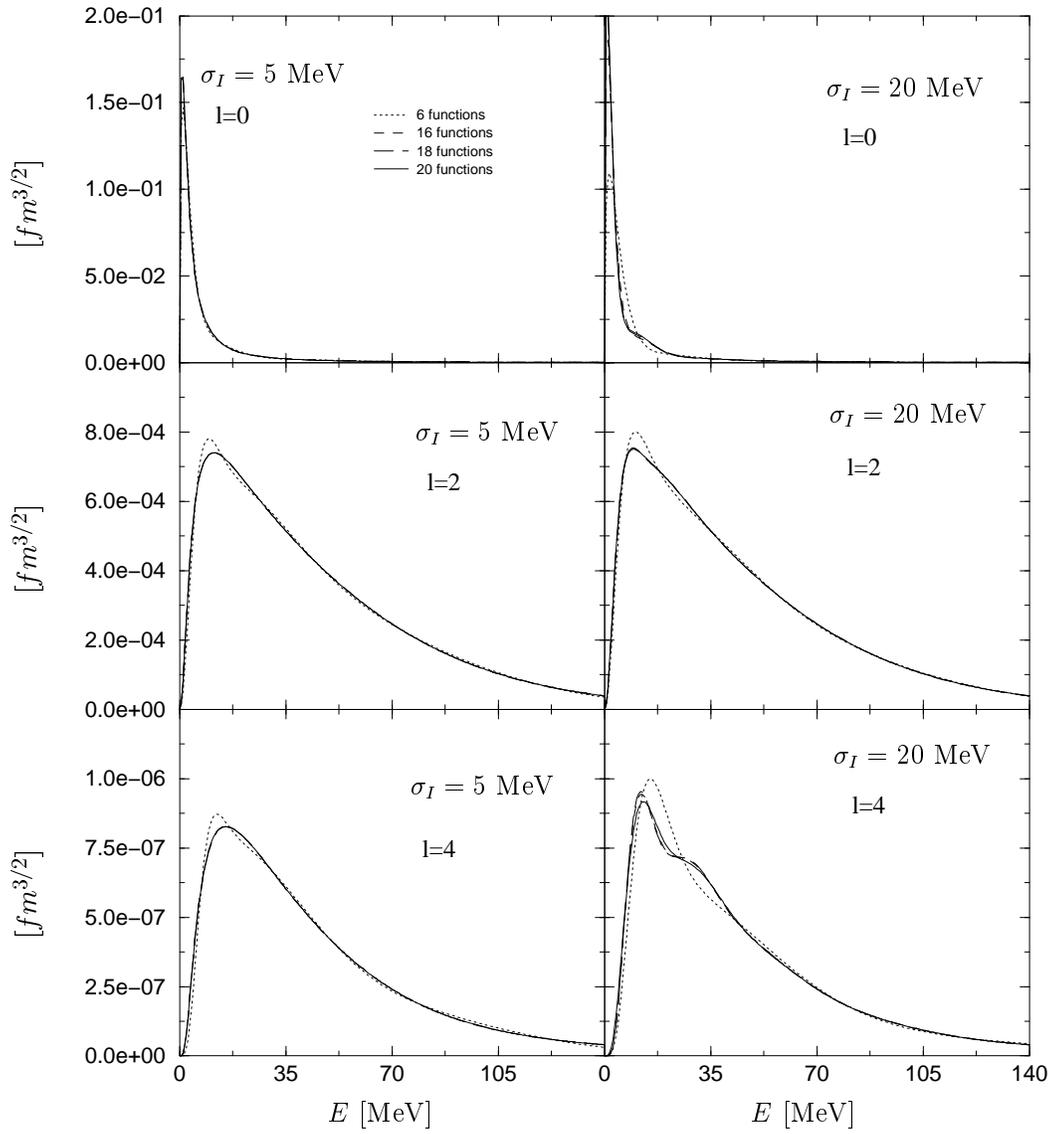
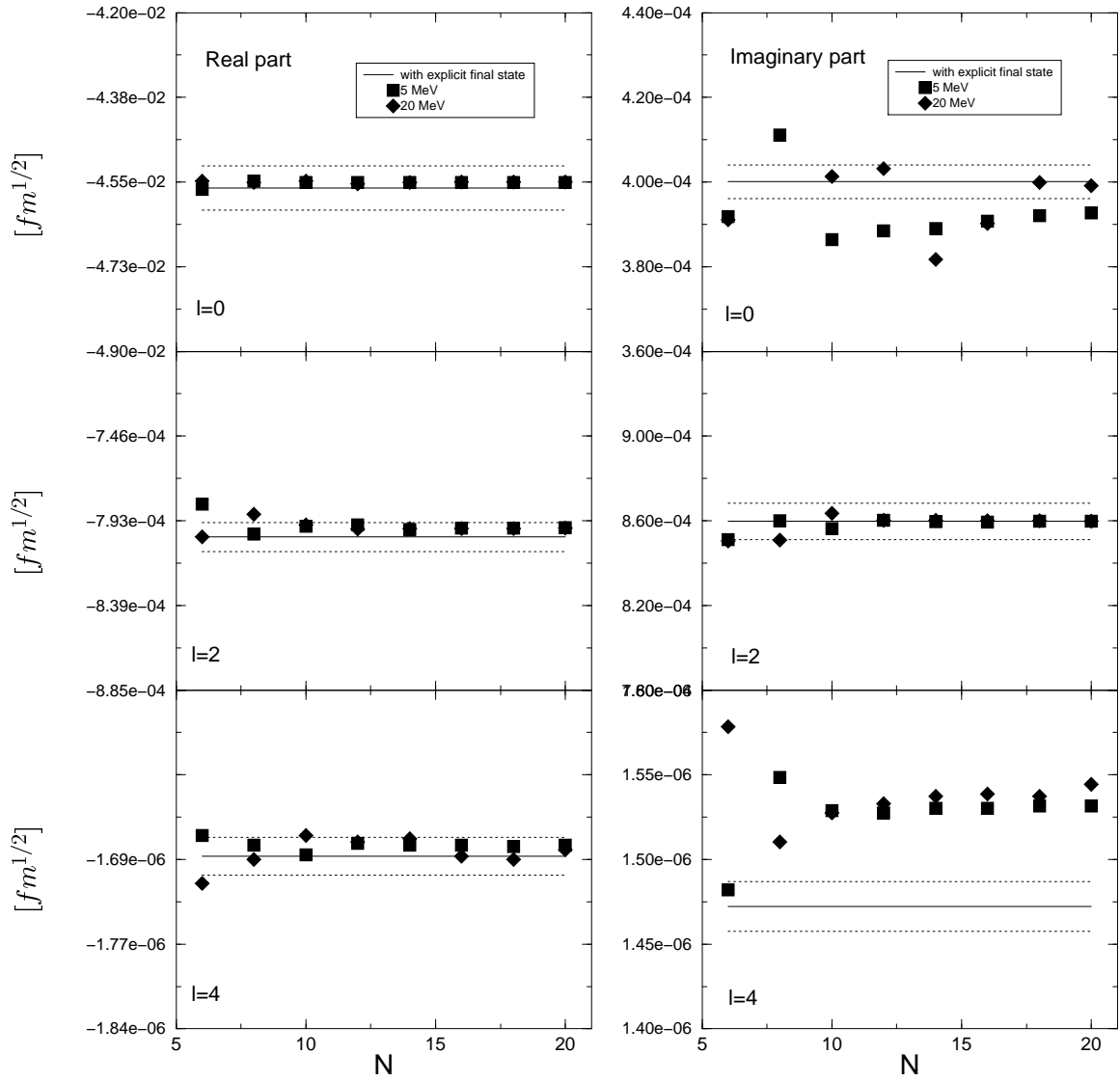
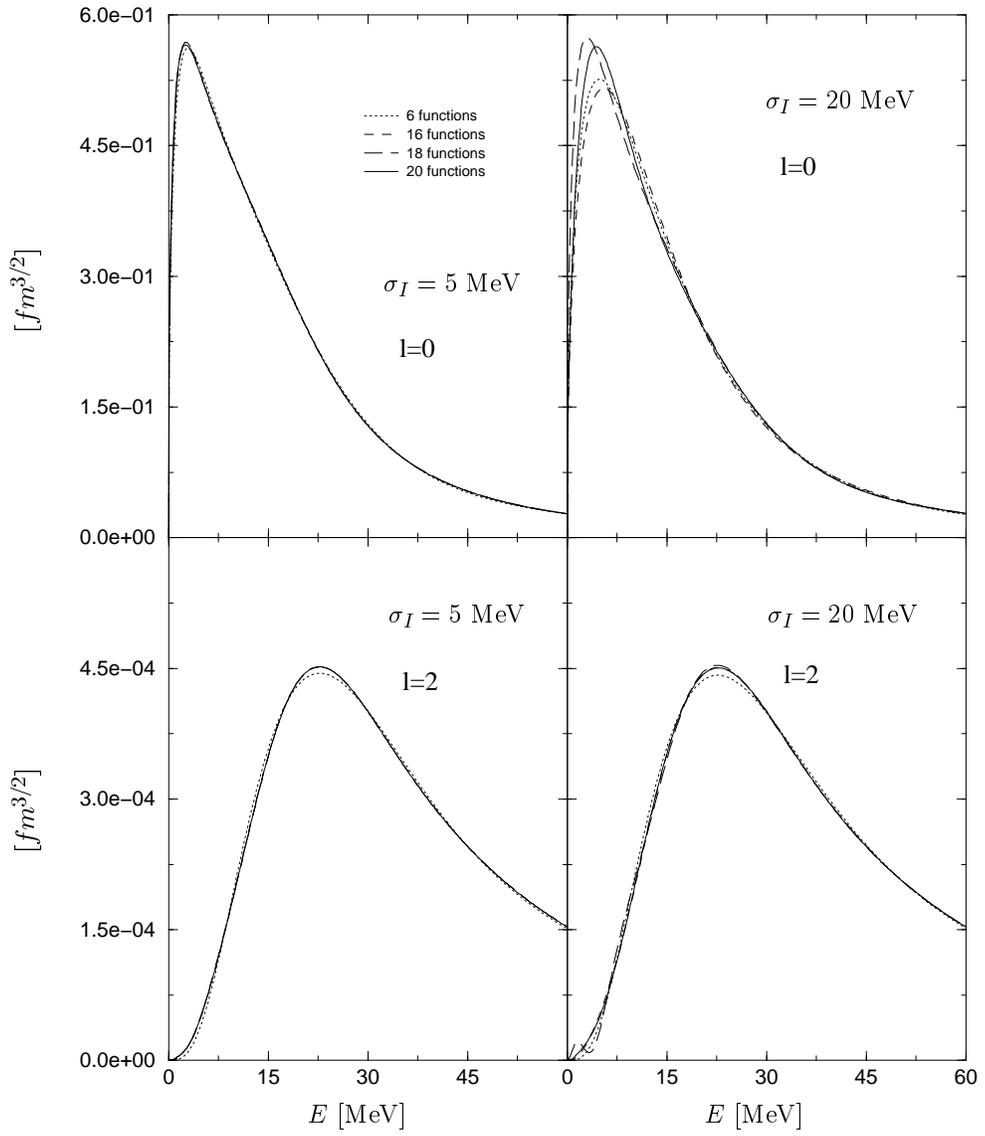
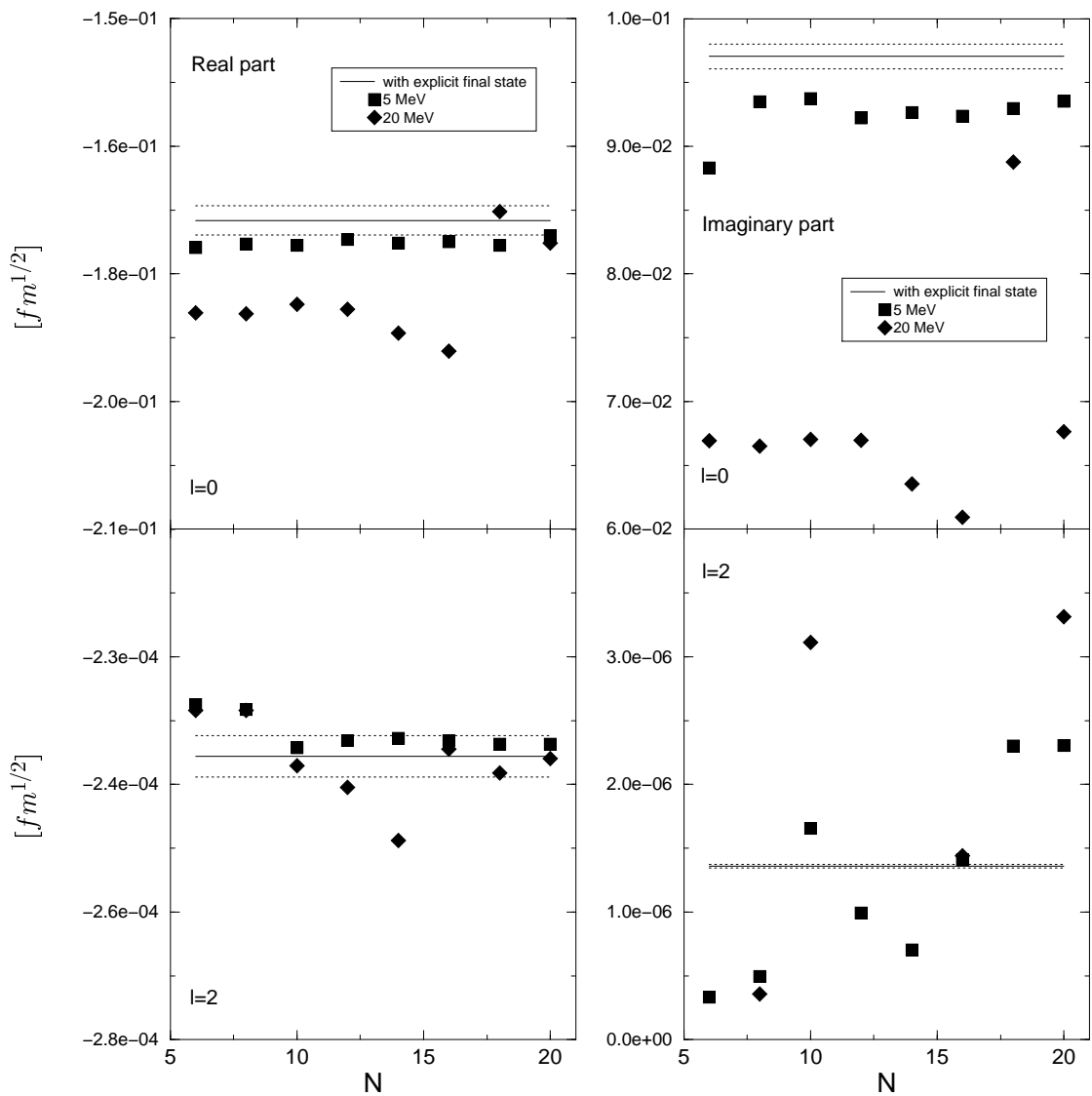
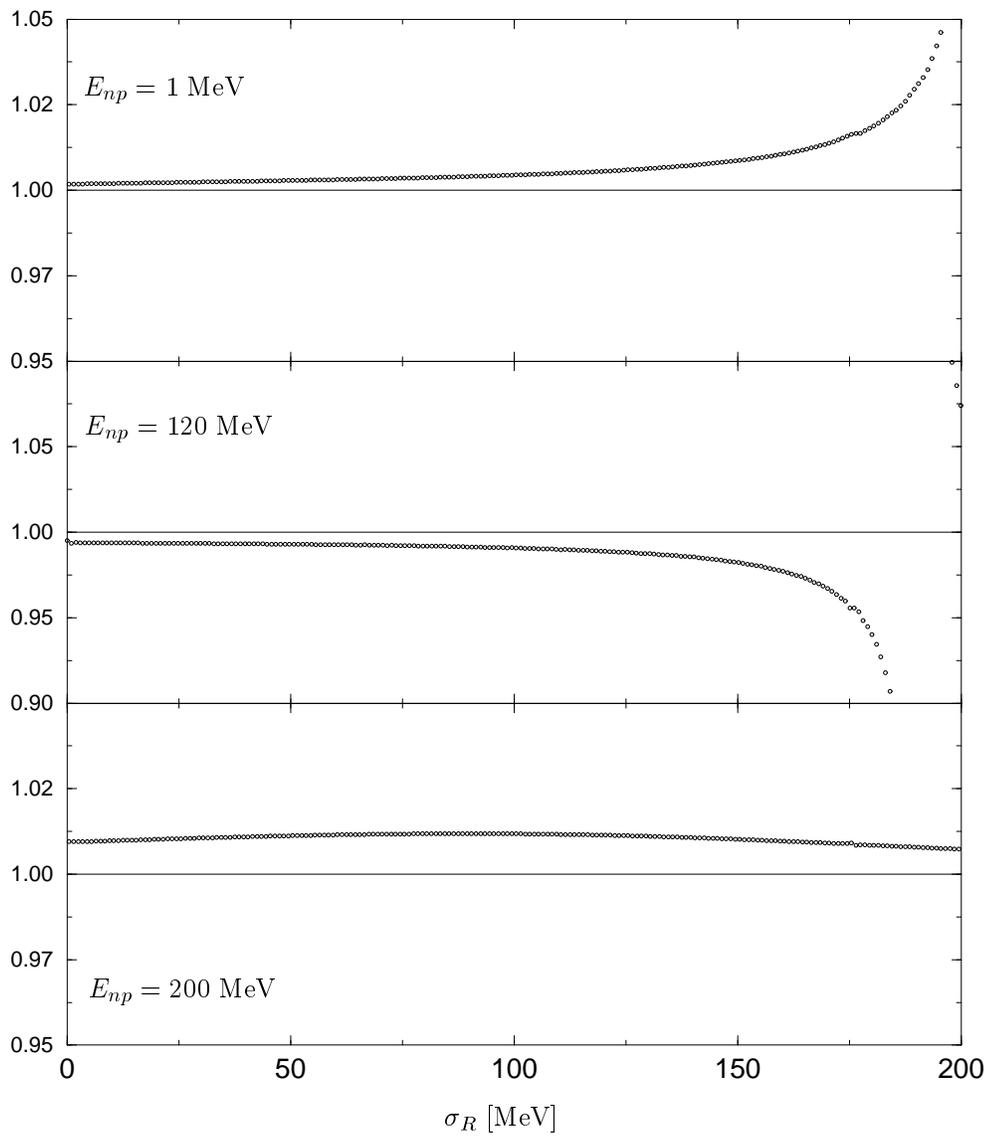
\end{document}